\documentclass[pra,twocolumn,superscriptaddress,aps,amsmath,amssymb,nofootinbib]{revtex4}
\usepackage{graphicx}
\usepackage{dcolumn}
\usepackage{bm}
\usepackage{color}

\usepackage[section]{placeins}
\usepackage{graphicx,epsfig}
\usepackage{amsmath}
\usepackage{amsfonts}
\usepackage{braket}
\usepackage{fancyhdr}
\usepackage{hyperref}
\usepackage[utf8]{inputenc}
\usepackage[T1]{fontenc}
\usepackage{amsthm}
\usepackage{hhline}
\usepackage{multirow}
\usepackage[figuresleft]{rotating}
\def\beq{\begin{equation}}
\def\eeq{\end{equation}}
\def\beqa{\begin{eqnarray}}
\def\eeqa{\end{eqnarray}}

\begin{document}



\title{Low-density neutron matter and the unitary limit}

\author{Isaac Vida\~na}
\affiliation{Istituto Nazionale di Fisica Nucleare, Sezione di Catania, Dipartimento di Fisica ``Ettore Majorana'', Universit\`a di Catania, Via Santa Sofia 64, I-95123 Catania, Italy}


\begin{abstract}

We review the properties of neutron matter in the low-density regime. In particular, we revise its ground state energy and the superfluid neutron pairing gap, and analyze their evolution from the weak to the strong coupling regime. The calculations of the energy and the pairing gap are performed, respectively, within  the Brueckner--Hartree--Fock approach of nuclear matter and the BCS theory using the chiral nucleon-nucleon interaction of Entem and Machleidt at N$^3$LO and the Argonne V18 phenomenological potential. Results for the energy are also shown for a simple Gaussian potential with a strength and range adjusted to reproduce the $^1S_0$ neutron-neutron scattering length and effective range. Our results are compared with those of quantum Monte Carlo calculations for neutron matter and cold atoms. The Tan contact parameter in neutron matter is also calculated finding a reasonable agreement with experimental data with ultra-cold atoms only at very low densities.  We find that low-density neutron matter exhibits a behavior close to that of a Fermi gas at the unitary limit, although, this limit is actually never reached. We also review the properties (energy, effective mass and quasiparticle residue) of a spin-down neutron impurity immersed in a low-density free Fermi gas of spin-up neutrons already studied by the author in a recent work where it was shown that these properties are very close to those of an attractive Fermi polaron in the unitary limit.

\end{abstract}

\maketitle


\section{Introduction}

Pure neutron matter \cite{Pethick95} is an ideal infinite nuclear system whose properties are of remarkable interest for a comprehensive understanding of neutron stars and neutron-rich nuclei. Particularly interesting are the properties of neutron matter at low densities, since they are crucial to understand the physics of the  inner crust of neutron stars \cite{Chamel08}, where the number density varies from $\sim 10^{-3}$ to $\sim 0.08$ fm$^{-3}$ and matter consists of a mixture of very neutron-rich nuclei (arranged in a Coulomb lattice), electrons and a superfluid neutron gas. Low-density neutron matter, however, is a system less trivial than one could expect at a first sight. The reason is that at low densities the neutron-neutron interaction is dominated by the $^1S_0$ partial wave which is very attractive and, although it is not able to bind two neutrons, leads to a well known virtual state which makes the neutron-neutron scattering length in this channel very large, $a_s=-18.9(4)$ fm \cite{Chen08}. Therefore, even at very low densities, where the average distance between two neutrons ($\propto k_F^{-1}$ with $k_F$ the Fermi momentum) is much larger than the effective range of the $^1S_0$ neutron-neutron interaction, $r_e=2.75(11)$ fm \cite{Miller90}, neutron matter is still a strongly correlated system.

Low-density neutron matter is similar to a unitary Fermi gas, an idealized system of spin-1/2 fermions with a zero-range interaction having an infinite (negative) scattering length in which all its properties are simply proportional to the corresponding ones of a non-interacting Fermi gas. 
The so-called unitary limit was introduced in 1999 by George Bertsch \cite{Bertsch01}, when he proposed a model of low-density neutron matter with a zero-range interaction and a tuned to infinity scattering length. In this limit, or close to it, all the length-scales of a system drop out and the Fermi momentum becomes the only relevant one. Dilute fermionic systems with $r_e\ll k_F^{-1}\ll|a_s|$, like neutron matter at low densities, exhibit universal properties close to those of a unitary Fermi gas, regardless the nature of the particles that constitute the system and their mutual interactions. Universality is expected to show in ground state properties \cite{Baker99}, collective excitations \cite{Stringari04,Heiselberg04,Kinast04,Kinast04b,Bulgac05,Altmeyer07,Wright07} and thermodynamical 
properties \cite{Kinast05,Thomas05,Bulgac05b,Bulgac06,Bulgac07}. In particular, the ground state energy of any fermionic system close to the unitary limit is expected to be $E=\xi\,E_{FG}$ where $\xi$ is the so-called Bertsch parameter and  $E_{FG}=3\hbar^2k_F^2/10m$ is the energy of the corresponding non-interacting Fermi gas. Different theoretical calculations predict values of $\xi$ in the range $0.3-0.7$ \cite{Baker99,Heiselberg01,Bruun04, Perali04,Nishida06,Haussmann07,Chen07}. The best estimations of the value of the Bertsch parameter come from quantum Monte Carlo (QMC) calculations which predict $\xi=0.44(1)$ \cite{Carlson03},  0.42(1) \cite{Astrakharchik04} and 0.40(1) \cite{Gezerlis08}. More recent QMC calculations from Carlson {\it et al.} predict, however, a slightly lower value $\xi=0.372(0.005)$ \cite{Carlson11}. Variational \cite{Friedman81}, finite volume Green's function Monte Carlo \cite{Carlson03b} and Brueckner--Hartree--Fock (BHF) \cite{Baldo08} calculations of the equation of state (EoS) of low-density neutron matter give $\xi\approx 0.5$. Using unitary nucleon potentials, constructed {\it ad hoc} to have an infinite $^1S_0$ neutron-neutron scattering length, the authors of Refs.\ \cite{Siu08,Dong10} studied the ground state energy of low-density neutron matter, obtaining values of $\xi$ remarkably close to the QMC predictions over a wide range of low densities, and showing, as expected, that low-density neutron matter behaves as a unitary Fermi gas as long as $a_s\rightarrow -\infty$.

Unitary Fermi gases  have been experimentally realized with ultra-cold trapped alkali atoms (with $^6$Li and $^{40}$K being the most commonly used ones), where the effective range of the interaction is $r_e\sim 10^{-4}k_F^{-1}$, and the scattering length $a_s$ can be tuned from negative to positive values with the help of magnetic fields, becoming infinity at the so-called Feshbach resonance \cite{Chin10}. These experiments provide constraints on the properties of unitary Fermi gases and, therefore, indirectly also on those of low-density neutron matter. Previous experimental measurements of the Berstch parameter with ultra-cold atomic gases reported the values $0.51(4)$ \cite{Kinast05}, $0.36(15)$ \cite{Bourdel04}, $0.46(5)$ \cite{Partridge06}, $0.46^{+0.05}_{-0.12}$ \cite{Stewart06}, and $0.39(2)$ to $0.435(15)$ \cite{Luo09}. The most precise experimental value of $\xi$ until now is $0.376(4)$ measured in 2012 by Ku {\it et al.} \cite{Ku12}. The possibility of varying in these experiments the interaction between the atomic species from a weak to a strong coupling regime in a controlled way, has additionally allowed the study of the whole crossover from BCS pairing with weakly attractive ($a_s<0$) Cooper pairs to the Bose--Einstein condensation (BEC) of bound dimers ($a_s>0$) \cite{Bloch08,Giorgini08,Calvanese18}. As it was mentioned before, although the neutron-neutron interaction is very attractive in the $^1S_0$ channel, it is unable to lead to the formation of a bound dineutron state and, hence, a BEC phase does not exist in neutron matter. Nonetheless, by varying the density, dilute neutron matter can go from the strong coupling regime close to the unitary limit to the weakly coupled BCS one. The importance in low-density neutron matter  of BCS-BEC crossover-like physics was pointed out by Matsuo in Ref.\ \cite{Matsuo06}, where he studied the behavior of the strong spatial dineutron correlation, finding that the density region region $n\approx (10^{-4}-0.5)n_0 $ (where $n_0\approx 0.16$ fm$^{-3}$ is the nuclear saturation density) corresponds to the domain of the BCS-BEC crossover. It is known from a general argument (see {\it e.g.,} Refs.\ \cite{Leggett80,Leggett80b,Nozieres85}), which applies to any dilute fermionic system, that the pair correlations of fermions interacting with a large scattering length differ from what is considered in the conventional BCS theory. Corrections due to pair correlations in the normal phase of neutron matter have been considered by several authors \cite{Ramanan13,Ramanan18,Tajima19,Urban20,Ohashi20,Inotani20} using the Nozi\`eres--Schmitt--Rink approach \cite{Nozieres85}, which is the simplest one that  interpolates correclty between the BCS and BCE limits. BCS-BEC crossover effects and the existence, above the critical temperature $T_c$ for the transition to the superfluid state, of a pseudo-gap in neutron matter have been also recently studied within the in-medium $T$-matrix formalism by Durel and Urban in Ref.\  \cite{Durel20}.  
At the BCS-BEC transition point, {\it i.e.} at the unitary limit, the superfluid pairing gap $\Delta$ is expected to be proportional to the free Fermi energy, $E_{F}=\hbar^2k_F^2/2m$. Ultra-cold atoms experiments with imbalance Fermi gases of $^6$Li found $\Delta=(0.45\pm0.05)E_F$ \cite{Shin08a,Carlson08,Schirotzek08}, in contrast with conventional superfluids or superconductors where the pairing gap is very weak, of the order of $\sim 0.1\%$ of the Fermi energy. QMC calculations of the neutron $^1S_0$ pairing gap by Gezerlis and Carlson \cite{Gezerlis10} found a maximum value of  $\Delta$  of  $\sim 0.3E_F$ at the Fermi momentum $k_F\sim 0.27$ fm$^{-1}$ ($n\sim 7\times 10^{-4}$ fm$^{-3}$). This maximum value of the gap corresponds to a strong coupling situation ($(k_Fa_s)^{-1}\sim -0.2$) close to that found 
in a unitary Fermi gas.

Experiments with population-imbalanced ultra-cold atomic gases, have allowed also to study the properties of polarized unitary  gases and quantum impurities leading, particularly, to the experimental realization of attractive and repulsive Fermi and Bose polarons, quasiparticles arising from the dressing of an impurity strongly coupled to a bath of particles of fermionic or bosonic nature. In the unitary limit, the energy of a Fermi polaron shows also a universal behavior, being $E_{pol}=\eta E_F$ \cite{Chevy06} with $\eta \approx -0.6$ \cite{Prokofev08,vanHoucke20,Shin08,Schirotzek09}. A few years ago, Forbes {\it et al.} \cite{Forbes14} extended the idea of the polaron to a system of strongly interacting neutrons and studied the energy of the neutron polaron with the QMC method. Similarly, Roggero {\it et al.} \cite{Roggero15} used also this method  to analyze the energy of a proton impurity in low-density neutron matter finding that, for a wide range of densities, the behavior of the proton impurity is similar to that of a polaron in a fully polarized unitary Fermi gas. Using the BHF approach,
very recently, in Ref.\ \cite{Vidana21} the author of the present work have analyzed the energy, effective mass and quasiparticle residue of a spin-down neutron impurity in a low-density free Fermi gas of spin-up neutrons, showing that these properties are in remarkable agreement with those of the attractive Fermi polaron in the unitary limit realized in ultra-acold atomic gases experiments.

In this work we review the properties of neutron matter in the low-density regime. Particularly, we revise its ground state energy and the superfluid neutron pairing gap, and analyze their evolution from the weak to the strong coupling regime. We use the well known BHF approach and the BCS theory to calculate, respectively, the ground state energy and the pairing gap, employing as bare nucleon-nucleon (NN) interactions the chiral one of Entem and Machleidt at N$^3$LO with a 500 MeV cut-off  (hereafter referred to simply as EM500)  \cite{Entem03} and the Argonne V18 (AV18) phenomenological potential \cite{Wiringa95}. The ground state energy is also calculated for a simple Gaussian potential with a strength and range adjusted to reproduce the $^1S_0$ neutron-neutron scattering length and effective range. Our results are compared with those of quantum Monte Carlo calculations for neutron matter and cold atoms.  Finally, we also review the properties  of a spin-down neutron impurity immersed in a low-density free Fermi gas of spin-up neutrons.

The manuscript is organized in the following way. The ground state energy of low-density neutron matter and the superfluid neutron pairing gap are presented, respectively, in Secs.\ \ref{sec:energy} and \ref{sec:pairing}, whereas, the properties of a spin-down neutron impurity in a low-density free Fermi gas of spin-up neutrons are shown in Sec.\ \ref{sec:polaron}. Finally, a brief summary and the main conclusions of this work are given in Sec.\ \ref{sec:conclusions}.

\begin{figure}[t]
\centering
\includegraphics[width=1.0 \columnwidth]{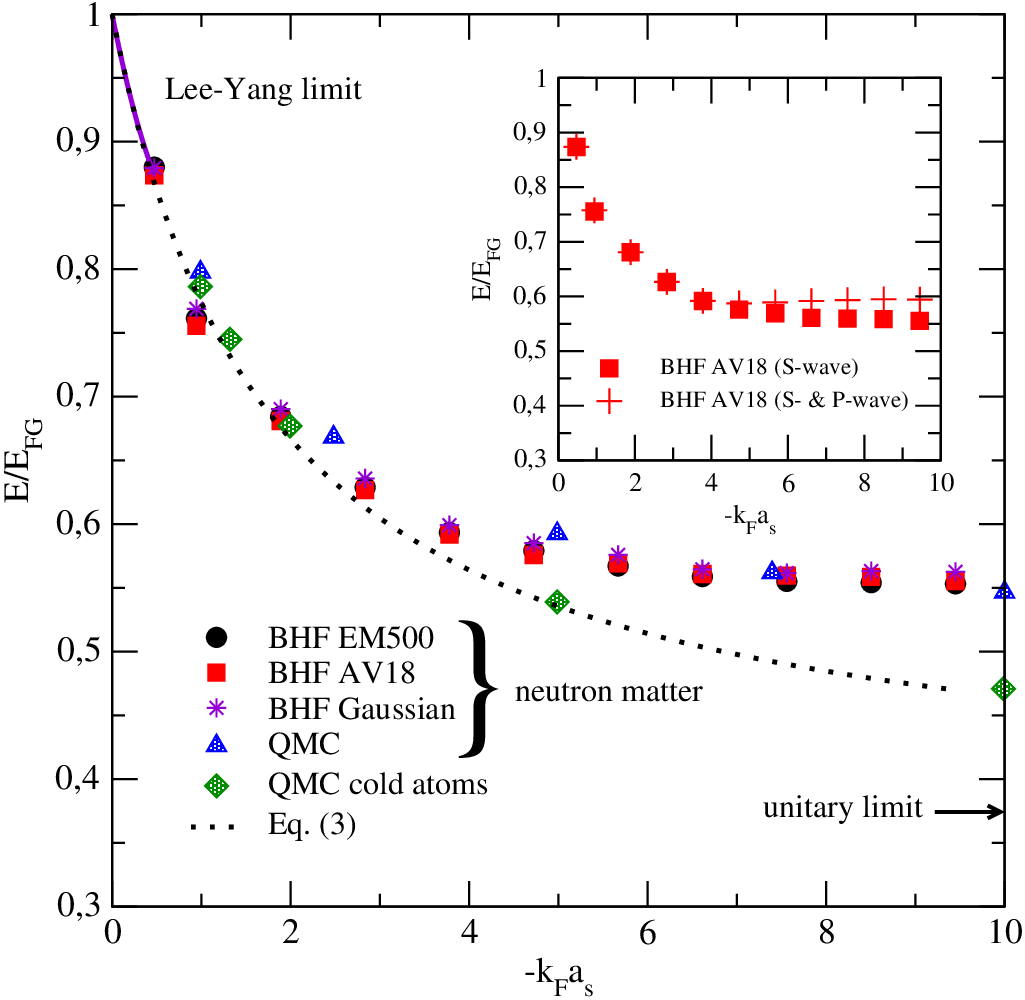}
        \caption{Ground state energy (in units of  $E_{FG}$) of low-density neutron matter as a function of the dimensionless parameter 
$-k_Fa_s$. Results are shown for our BHF calculation (full circles, full squares and stars) and the QMC one (full triangles) of Ref.\ \cite{Gezerlis08}. The energy of cold atoms (full diamonds), obtained also by the authors of Ref.\  \cite{Gezerlis08}, is shown for comparison. The continuous line displays the Lee--Yang limit (see Eq.\ (\ref{e:leeyang})) for $-k_Fa_s\ll 1$ \cite{Lee57}. The arrow indicates the most precise experimental value $\xi=0.376(4)$ \cite{Ku12} of the Bertsch parameter measured with ultra-cold atomic gases at the unitary limit. The dotted line shows the density functional proposed by Lacroix (see Eq.\ (\ref{e:lacroix})) in Ref.\ \cite{Lacroix16}. The 
inset shows, as illustration, the effect on the ground state energy of the inclusion of the $^3P_0, ^3P_1$ and $^3P_2$ partial waves in the case of the BHF calculation with the AV18 potential.} 
\label{fig:fig11}
\end{figure}

\section{Ground state energy}
\label{sec:energy}

We start this section by showing in Fig.\ \ref{fig:fig11} the ground state energy (in units of  $E_{FG}$) of low-density neutron matter and cold atoms as a function of the dimensionless parameter $-k_Fa_s$.  We note that, although the neutron matter results depend only on the Fermi momentum  because the value of $a_s$ is fixed, we use the product $k_Fa_s$ as independent variable to facilitate the comparison with the cold atom results which are usually presented as a function of it. Full circles and squares display the results for neutron matter of our BHF calculation performed using the EM500 interaction (full circles)  and the AV18 potential (full squares). BHF results are also presented here for a simple Gaussian NN potential (stars)
\begin{equation}
V(r)=V_0\,\mbox{e}^{-(r/r_0)^2} 
\label{eq:gauss}
\end{equation}
with a strength $V_0=-31.02215$ MeV and a range $r_0=1.8$ fm adjusted to reproduce the $^1S_0$ neutron-neutron scattering length and effective range. Only contributions from the  $^1S_0$ partial wave, which is the dominant one in the low-density region considered, are included in these calculations.  Contributions from three-nucleon forces are expected to be irrelevant at these densities and, therefore, are neglected in our calculations \cite{Hebeler10, Tews13,Kruger13}. Full triangles and diamonds correspond, respectively, to the QMC results for neutron matter (full triangles) and cold atoms (full diamonds) obtained by Gezerlis and Carlson in Ref.\ \cite{Gezerlis08}. The QMC results of neutron matter shown here include, as our BHF calculation, contributions only from the $^1S_0$ partial wave and were obtained also with the AV18 potential. To illustrate the effect of  P-wave interactions on the ground state energy we show in the inset of the figure, as an example, the result obtained for the BHF calculation with the AV18 potential when the $^3P_0, ^3P_1$ and $^3P_2$ partial waves are included, and compare it with the energy obtained when only the $^1S_0$
channel is considered. We find that the energy of neutron matter increases by $\sim 7\%$ at $-k_Fa_s=10$ while this increase is only about $(2-3)\%$ at $-k_Fa_s=5$,
and that the effect of the P-waves is completely negligible at lower densities. Similar results were found by Gezerlis and Carlson in Ref.\  \cite{Gezerlis10} using the Argonne V4 potential \cite{Wiringa02} (see figure 3 of Ref.\ \cite{Gezerlis10}). For the cold atom case, Gezerlis and Carlson considered an hyperbolic cosinus interaction potential of the form 
\begin{equation}
v(r)=-v_0\frac{2\hbar^2}{m}\frac{\mu^2}{\mbox{cosh}^2(\mu r)} \ , 
\label{e:cosh}
\end{equation}
where the strength $v_0$ was adjusted to obtained values of $-k_Fa_s$ from $1$ to $10$, and $\mu$ was taken such that the effective range of the potential was much smaller than the interatomic 
distance. 

Coming back to the figure, the arrow indicates the most precise experimental value, $\xi=0.376(4)$ \cite{Ku12}, of the Bertsch parameter measured with ultra-cold atomic gases
at the unitary limit ({\it i.e.,} for $-k_Fa_s\rightarrow \infty$), whereas the continuous line shows the well known extreme low-density limit ($-k_Fa_s\ll 1$) of Lee and Yang  \cite{Lee57}
\begin{equation}
\frac{E}{E_{FG}}=1+\frac{10}{9\pi}k_Fa_s+\frac{4}{21\pi^2}(11-2\mbox{ln}2)(k_Fa_s)^2 \ .
 \label{e:leeyang}
\end{equation}
In the figure, for comparison, we also show the recent density functional proposed by Lacroix in Ref. \cite{Lacroix16},
\begin{equation}
\frac{E}{E_{FG}}=1+\frac{10}{9\pi}\frac{k_Fa_s}{1-\frac{10}{9\pi}k_Fa_s/[1-\xi(k_Fr_e)]}
\label{e:lacroix}
\end{equation}
with
\begin{equation}
\xi(k_Fr_e)=1-\frac{(1-\xi_0)^2}{1-\xi_0+k_Fr_e\eta_e}
\label{e:lacroix2}
\end{equation}
where the two parameters $\xi_0$ and $\eta_e$ of this functional are fixed to reproduce both the universal properties of a unitary Fermi gas, and the 
Lee--Yang limit at extremely low densities. In particular, we show the Lacroix's functional for $\xi_0=0.37$ assuming that $r_e=0$, therefore, being the r. h.s. of Eq.\ (\ref{e:lacroix2}) simply reduced to $\xi_0$, which makes the value of $\eta_e$ irrelevant in this case. 

We note first that, for neutron matter, our BHF calculation gives results very similar for the three NN interactions employed. Furthermore, we notice also that our results are in quite good agreement with the QMC ones over the whole range of values of the dimensionless parameter $-k_Fa_s$ considered. This indicates that not only the details of the NN interaction ({\it e.g.,} the value of the cut-off in the case of the chiral forces) are irrelevant for neutron matter a very low densities, but also those of the approach employed to solve the many-body problem seem to be quite unimportant (see, {\it e.g.}, figure 4 of Ref.\ \cite{Gezerlis10}, where it is shown that results for neutron matter obtained with different methods agree within $20\%$ in the range $0<-k_Fa_s<20$). It can be also seen in the figure that our BHF results as well as the QMC ones extrapolate properly to the Lee--Yang limit at extremely low densities, and that all them are reasonably well reproduced by the Lacroix's functional of Eq.\ (\ref{e:lacroix}) for $-k_Fa_s\leq2$. 

In the unitary limit QMC results of the ratio $E/E_{FG}$, in the case of cold atoms,  approach the value $0.37$ (see Ref.\  \cite{Carlson11}), in very good agreement with the most precise experimental measurement of the Berstch parameter \cite{Ku12}.  As we already said, neutron matter never reaches strictly the unitary limit, although is close to it. In particular, Baldo and Maieron \cite{Baldo08}, on the basis of the Brueckner--Bethe--Goldstone many-body theory,  showed that in the range of densities corresponding to the Fermi momenta $0.4<k_F<0.8 $
fm$^{-1}$ the energy of neutron matter turns to be very close to one half of $E_{FG}$. A similar result was found also in the variational and 
finite volume Green's function Monte Carlo calculations of Refs.\ \cite{Friedman81} and \cite{Carlson03b}.  As it is seen in the figure, our BHF results show an almost constant value of the ratio  $E/E_{FG}$ in the range $6.61<-k_Fa_s<9.45$ ($0.35<k_F<0.5$ fm$^{-1}$). Making a linear fit of our results in this range we obtain, respectively, the values $E/E_{FG}=0.561$, $0.563$ and $0.566$ for the EM500 interaction, the AV18 potential and the Gaussian potential, in agreement with the results these works. The linear dependence of the neutron matter energy with $E_{FG}$, found in our BHF calculation in this Fermi momentum range, can be understand from an argument pointed out by Carlson {\it et al.} in Ref.\ \cite{Carlson03b} that we briefly review here.

\begin{figure}[t]
\centering
\includegraphics[width=1.0 \columnwidth]{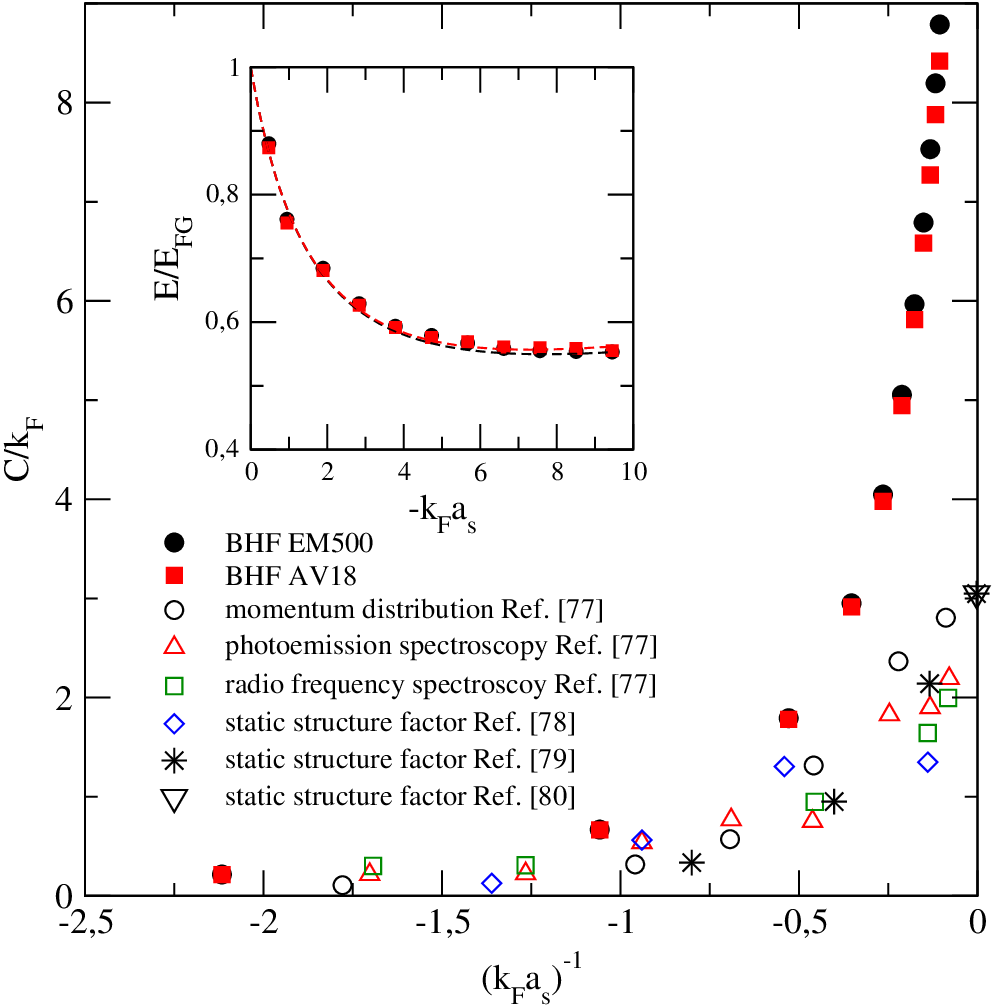}
        \caption{Tan contact parameter as a function of $(k_Fa_s)^{-1}$ in units of the Fermi momentum. Results for neutron matter from our BHF calculation obtained with the EM500 interaction (full circles) and the AV18 potential (full squares) are compared with data from experiments with ultra-cold fermionic gases of $^{40}$K (open circles, triangles and squares) \cite{Stewart10} and $^6$Li (open diamonds, stars and open down triangle) \cite{Kuhnle10,Kuhnle11,Hoinka13}. The inset shows the fit of our BHF results for the ground state energy of neutron matter with the Lacroix's functional of Eq.\ (\ref{e:lacroix}) (dashed lines).} 
\label{fig:fig12}
\end{figure}

The interaction energy is proportional to the density ($\propto k_F^3$) times the volume integral of the $G$-matrix which is related to the bare interaction $V$ through the well known Brueckner equation, written schematically as
\begin{equation}
G\phi=V\psi \ ,
\label{e:bge}
\end{equation}
where $\phi$ and $\psi$ are the unperturbed and perturbed two-neutron wave functions. At low densities all the relevant relative momenta are small and, therefore, one has $\phi=1$ and, in vacuum, beyond the effective range of the interaction, $\psi= 1-a_s/r_e$. In addition, since for neutron matter one has $-a_s/r_e > 1$ one can approximate $\psi$ simply by $-a_s/r_e$, and, therefore, $G\approx -a_sV/r_e$. It is easy to see then that in the range $6.61<-k_Fa_s<9.45$, the $G$-matrix is proportional to $(k_Fr_e)^{-1}$. Consequently, its volume integral becomes, in this range, proportional to $k_F^{-1}$ and the interaction energy proportional to $k_F^2$, as it is the case of the energy of the non-interacting Fermi gas. From now on, BHF results will be presented only for the EM500 interaction and the AV18 potential.

To finish this section we show in Fig.\ \ref{fig:fig12} our results for the Tan contact parameter \cite{Tan08a,Tan08b,Tan08c} of neutron matter as a function of $(k_Fa_s)^{-1}$. The contact parameter of an infinite spin saturated system is given by
\begin{equation}
C=\frac{1}{n}\frac{4\pi ma_s^2}{(\hbar c)^2}\frac{d\varepsilon}{da_s},
\label{e:tan}
\end{equation}
where $n=k_F^3/3\pi^2$ and $\varepsilon=nE$ are, respectively, the density and the energy density of the system. For computational reasons, to calculate $C$ we have first fitted our BHF results for the ground state energy $E$ of neutron matter by using the Lacroix's functional of Eq.\ (\ref{e:lacroix}) taking $r_e=2.75$ fm. A good fit of our results for the energy is obtained using the parameters $\xi_0=0.326$ and $\eta_0=0.15$ in the case of the EM500 interaction, and $\xi_0=0.326$ and $\eta=0.165$ for the AV18 potential. The results of the fit for are shown by the dashed lines in the inset of the figure. Our result for the contact parameter in neutron matter is compared with the experimental data from measurements with ultra-cold fermionic atoms of the momentum distribution (open circles), photoemission spectroscopy (open triangles) and radio frequency spectroscopy (open squares) of a gas of $^{40}$K \cite{Stewart10}, and the static structure factor (open diamonds, stars and open down triangle ) of a gas of $^6$Li \cite{Kuhnle10, Kuhnle11,Hoinka13}. A reasonable good agreement between our results for neutron matter and the experimental data from ultra-cold atoms is found only in the very low-density regime ($(k_Fa_s)^{-1}<-1$), being the differences very large in the range $-0.15 < (k_Fa_s)^{-1}<-0.10$, near the unitarity limit. The reason for these large differences is probably the fact that low-density neutron matter, as it has been already said, actually never reaches the unitary limit. 


\begin{figure}[t]
\centering
\includegraphics[width=1.0 \columnwidth]{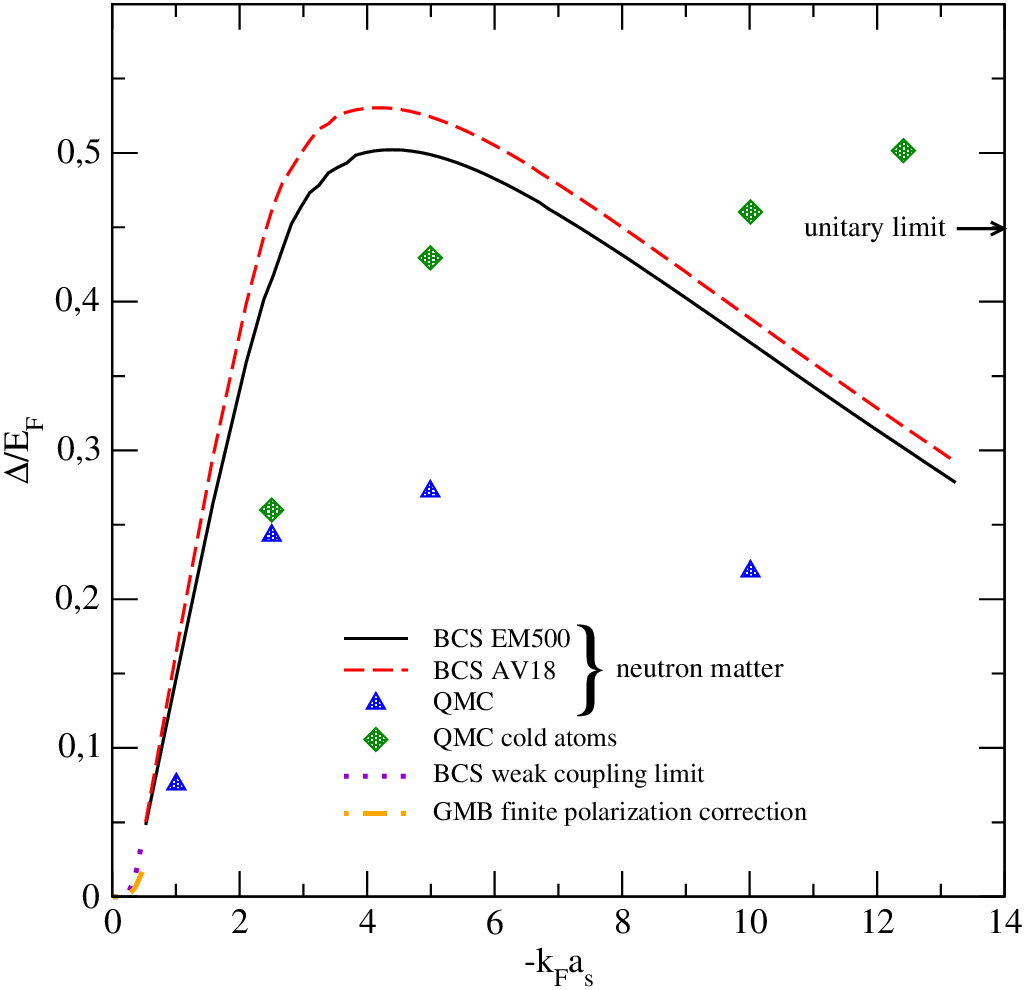}
        \caption{Superfluid $^1S_0$ pairing gap (in units of  $E_{F}$) of low-density neutron matter as a function of the dimensionless parameter 
$-k_Fa_s$.  BCS results obtained with the EM500 interaction and the AV18 potential are shown together with the QMC ones (full triangles) of Ref.\ \cite{Gezerlis08}. The superfluid pairing gap for cold atoms (full diamonds), calculated also by the authors of Ref. \cite{Gezerlis08}, is shown for comparison. The dotted and dot-dashed lines at very low values of $-k_Fa_s$  display, respectively, the BCS result in the weak coupling limit and the Gorkov and Melik-Barkhudarov \cite{Gorkov61} finite polarization correction. The arrow indicates the value of $\Delta/E_F$ at the unitarity limit extracted from ultra-cold Fermi atoms experiments with imbalance mixtures of $^6$Li \cite{Shin08a,Carlson08,Schirotzek08}.} 
\label{fig:fig21}
\end{figure}

\section{Superfluid Pairing Gap}
\label{sec:pairing}

We consider now the superfluid pairing gap of neutron matter at low densities, which is an important quantity to understand the properties of neutron-rich nuclei \cite{Liutvinov05} and neutron star cooling \cite{Fortin10}. In particular, we have performed a mean-field BCS calculation of the $^1S_0$ pairing gap using the EM500 interaction and the AV18 potential. The results of this calculation are shown (in units of $E_F$) as a function of the dimensionless parameter $-k_Fa_s$  in Fig.\ \ref{fig:fig21}, together with those of the QMC ones obtained by Gezerlis and Carlson (full triangles) \cite{Gezerlis08} using also the AV18 potential. The superfluid pairing gap for cold atoms (full diamonds) \cite{Gezerlis08}, calculated also by these two authors with the hyperbolic cosinus interaction of Eq.\ (\ref{e:cosh}), is shown for comparison. The dotted and dot-dashed lines at very low values of the parameter $-k_Fa_s$ show, respectively, the well known analytic  BCS result in the weak coupling limit  
\begin{equation}
\Delta^0_{BCS}(k_F)=\frac{8}{e^2}\frac{\hbar^2k_F^2}{2m}\mbox{exp}\left(\frac{\pi}{2k_Fa_s}\right) \ ,
\label{e:bcsweak}
\end{equation}
where $e$ is the Euler's number, and the Gorkov and Melik--Barkhudarov (GMB) \cite{Gorkov61} finite polarization correction,
\begin{equation}
\Delta^0(k_F)=\frac{1}{(4e)^{1/3}}\frac{8}{e^2}\frac{\hbar^2k_F^2}{2m}\mbox{exp}\left(\frac{\pi}{2k_Fa_s}\right)\ , 
\label{e:gmb}
\end{equation}
due to the inclusion of induced interactions that reduce the gap even at weak coupling. As it is seen, QMC results for both neutron matter and cold atoms extrapolate properly to the GMB result whereas our BCS calculation, as expected,  does it towards the BCS weak coupling limit.

\begin{figure*}[t!]
\centering
\includegraphics[width=1.5 \columnwidth]{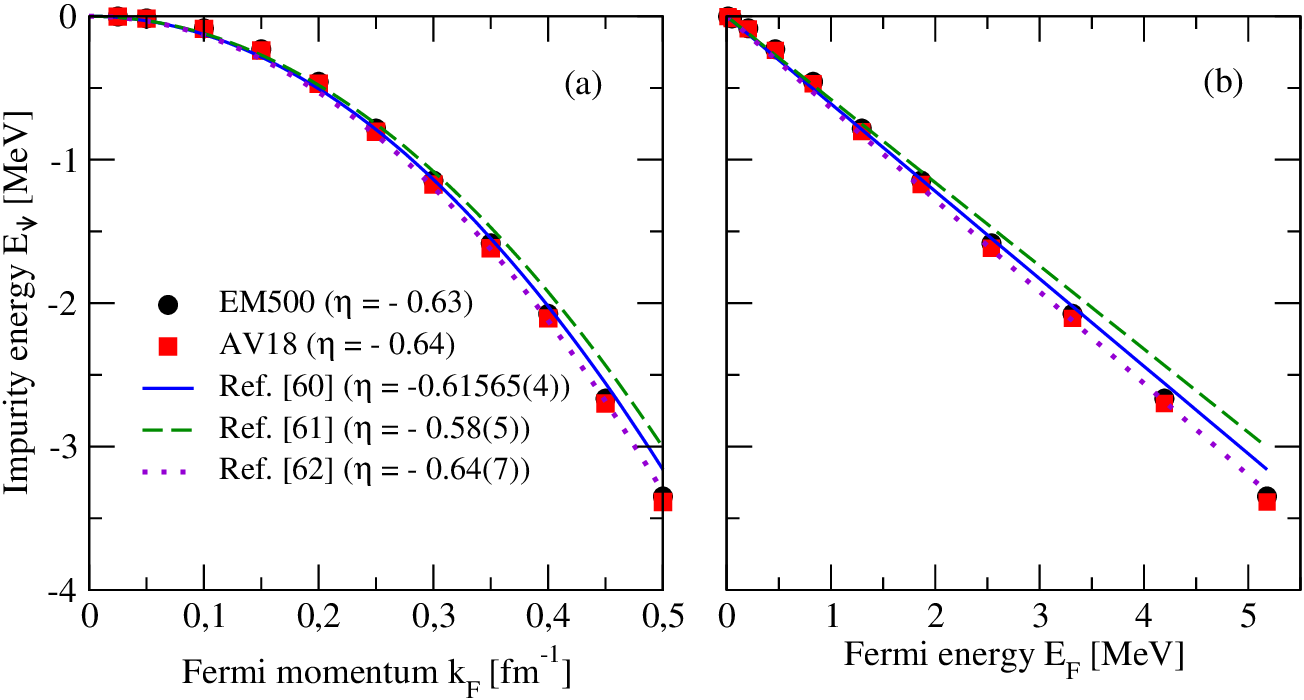}
        \caption{Energy of a $\downarrow$ neutron impurity with zero momentum as a function of the Fermi momentum (panel a) and of the Fermi energy (panel b) of the free gas of $\uparrow$ neutrons. The results obtained with the EM500 interaction (full circles) and the AV18 potential (full squares) are compared with those obtained when using the values of the proportionality constant $\eta$ derived  in the QMC calculation of Ref.\ \cite{vanHoucke20} (solid line) and experimentally in Refs.\ \cite{Shin08,Schirotzek09} (dashed and dotted lines).}
\label{fig:fig41}
\end{figure*}

We already said in the introduction that close to the unitary limit the superfluid pairing gap is expected to be  $\Delta=\delta E_F$, where the proportionality constant $\delta$ was found $\sim 0.45$ in ultra-cold Fermi atoms experiments with imbalance mixtures of $^6$Li  \cite{Shin08a,Carlson08,Schirotzek08}. This value is indicated in the figure with an arrow. At the unitary limit, the QMC result for cold atoms of Gezerlis and Carlson is $0.50(3)E_F$, in good agreement with these experiments. For neutron matter, the QMC calculation predicts, as we also mentioned in the introduction, a maximum value of the pairing gap of $\sim 0.3E_F$ at  $-k_Fa_s\sim 5$, which is $\sim 65\%$ of the value of the BCS result found by the same authors in Ref.\ \cite{Gezerlis08}. Our BCS calculation predicts a maximum value of $\Delta$ of $\sim 0.50E_F$ at $-k_Fa_s=4.4$ and of $\sim 0.53E_F$ at $-k_Fa_s\sim 4.1$ when using the EM500 interaction or the AV18 potential, respectively. It is interesting to note that while for the ground state energy both NN interactions give very similar results, this is not the case for the pairing gap for which their predictions are slightly different. The maximum values of the gap found in both QMC and BCS calculations correspond to a strong coupling situation where the behavior of neutron matter, with a Fermi momentum of  $\sim 0.27$ fm$^{-1}$ in the QMC case or $\sim 0.21-0.23$ fm$^{-1}$ in the BCS one, can be considered close to that of a unitary Fermi gas. The reader should note, however, that although the maximum value of the gap obtained with our BCS calculation seems to be in agreement with experimental data from ultra-cold atoms experiments, this is not the case because the BCS is just a mean-field calculation which does not include the effects of medium polarization that are very important even in the low-density regime and reduce the value of the gap. Therefore, in the case of our BCS calculation, our previous statement regarding the vicinity of neutron matter to the unitary limit should be considered only qualitatively.


\section{Neutron polaron}
\label{sec:polaron}

In this final section, we review the recent analysis of the energy, effective mass and quasiparticle residue of a spin-down 
($\downarrow$) neutron impurity immersed in a low-density free Fermi gas of spin-up ($\uparrow$) neutrons, made by the author of the present work in Ref.\ \cite{Vidana21} using the BHF approach, where he showed that the $\downarrow$ neutron impurity behaves basically as an attractive Fermi polaron in a 
unitary gas.

\begin{figure}[t]
\centering
\includegraphics[width=1.0 \columnwidth]{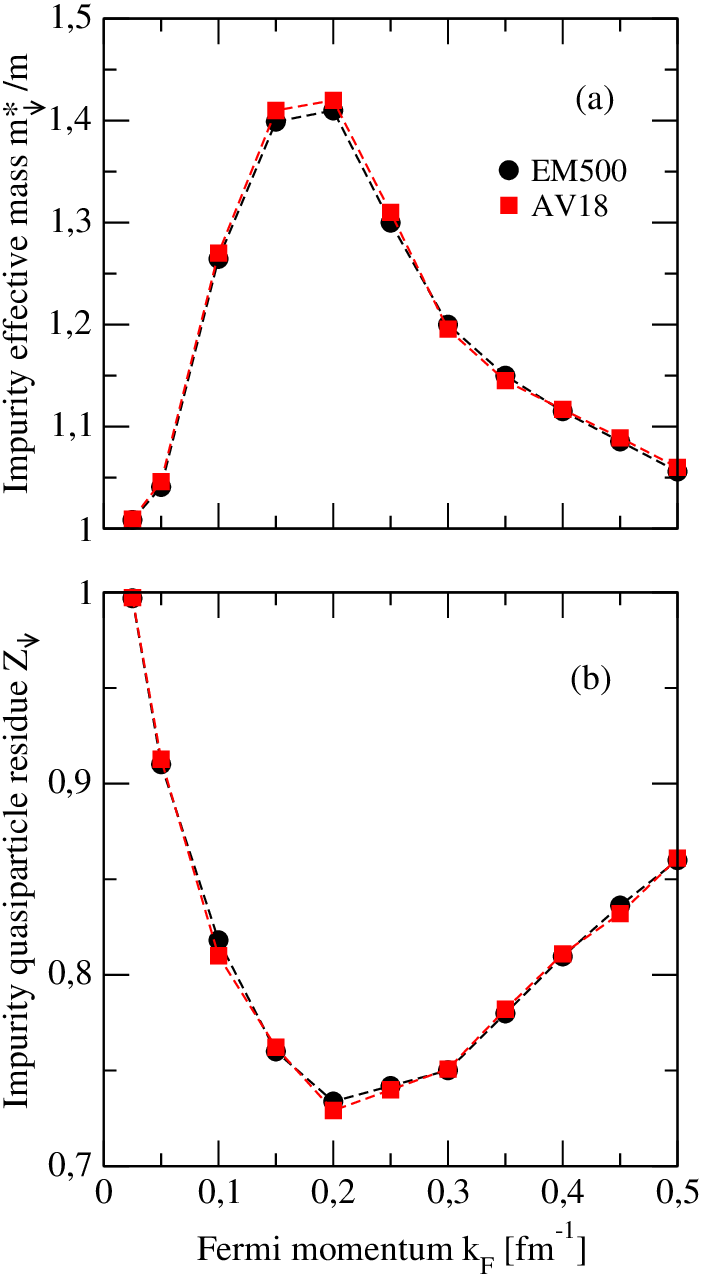}
        \caption{Effective mass (panel a) and quasiparticle residue (panel b) of a $\downarrow$ neutron impurity with zero momentum as a function of the Fermi momentum of the free gas of $\uparrow$ neutrons. Results are shown for the EM500 interaction (full circles) and the AV18 potential (full squares).}
\label{fig:fig42}
\end{figure}

The energy of the $\downarrow$ neutron impurity with zero momentum is shown in Fig.\ \ref{fig:fig41} as a function of the Fermi momentum (panel a) and of the Fermi energy (panel b) of the free gas of $\uparrow$ neutrons for the EM500 interaction (full circles) and the AV18 potential (full squares). Note that both interactions predict essentially the same results. Note in addition that the linear behavior shown by the energy of a Fermi polaron in the unitary limit, $E_{pol}=\eta E_F$ \cite{Chevy06}, is clearly seen also in the case of the energy the $\downarrow$ neutron impurity, where a value of the proportionality constant $\eta=-0.63$ ($\eta=-0.64$) is found using the EM500 interaction (AV18 potential). These numbers are in very good agreement with the results of  state-of-the-art QMC calculations $\eta=-0.61565(4)$ \cite{vanHoucke20} and the values $\eta=-0.58(5)$ \cite{Shin08} and $\eta=-0.64(7)$ \cite{Schirotzek09} extracted from experiments with spin-polarized  $^6$Li atoms with resonant interactions. Our result shows that a $\downarrow$ neutron impurity in a low-density free Fermi gas of $\uparrow$ neutrons presents a behavior similar to that of attractive Fermi polaron in the unitary limit, being irrelevant the details of the interaction between the impurity and the free Fermi gas. To further confirm this behavior, in the next, we analyze also the effective mass and the quasiparticle residue of a  $\downarrow$ neutron impurity with zero momentum.

The effective mass of a $\downarrow$ neutron impurity with zero momentum, $m^*_\downarrow$, can be extracted by assuming that its energy is quadratic for  low values of its momentum $\vec k_{\downarrow}$, and fitting this parabolic energy  to the energy calculated within the BHF approach. The quasiparticle residue is defined as
\begin{equation}
Z_\downarrow=\left(1-\frac{\partial U_\downarrow(\vec k_{\downarrow}=\vec 0,E'_{\downarrow})}{\partial E'_{\downarrow}}\right)^{-1}_{E'_\downarrow=U_\downarrow(\vec k_{\downarrow}=\vec 0)} 
\label{eq:zfact}
\end{equation}
where $U_\downarrow(\vec k_{\downarrow},E'_{\downarrow})$ is the off-shell BHF $\downarrow$ neutron potential. It gives a measurement of the importance of the correlations. The more important the correlations are, the smaller is its value. Results for both quantities are shown in panels a and b of Fig.\ \ref{fig:fig42} as a function of the Fermi momentum of the $\uparrow$ neutron free Fermi gas for the EM500 interaction (full circles) and the AV18 potential (full squares). Note that also in this case both interactions predict almost the same results, indicating once more the irrelevance of the interaction details in the low-density regime. As it can be seen in the figure, initially  $m^*_\downarrow$ ($Z_\downarrow$) increases (decreases), then it reaches a maximum (minimum) at 
$k_F\sim 0.2$ fm$^{-1}$ and finally it decreases (increases) at higher densities.  We notice that for $k_F\sim 0.2$ fm$^{-1}$, where $m^*_{\downarrow}$ and $Z_\downarrow$ present their respective maximum and minimum, the average interparticle spacing $n^{-1/3}$ (with $n=k_F^3/6\pi^2$ the density of the $\uparrow$ neutron free Fermi gas) is of the order of the $^1S_0$ neutron-neutron scattering length, {\it i.e.,} $n^{1/3}|a_s|\sim 1$. We can venture to say that $k_F\sim 0.2$ fm$^{-1}$ establishes the border between a less correlated and a more correlated regime of the system. In fact, note that  the values of  $Z_\downarrow$ are in general larger in the $k_F$ region from $0$ to $0.2$ fm$^{-1}$ than for $k_F \gtrsim 0.2$ fm$^{-1}$, indicating that in this region of very low-densities correlations are less important, and that the $\downarrow$ neutron impurity propagates more freely in the $\uparrow$ neutron gas. We also notice that for Fermi momenta above $\sim0.2$ fm$^{-1}$, the $^1S_0$ neutron-neutron scattering length is larger that the average interparticle spacing with values of the dimensionless quantity $n^{1/3}|a_s|$  ranging from $1$ at $k_F=0.21$ fm$^{-1}$ to $2.37$ at $k_F=0.5$ fm$^{-1}$. Although in the unitary limit it is strictly fulfilled the condition $n^{1/3}|a_s|\gg 1$, these numbers indicate once more that low-density neutron matter 
is close to it, at least for Fermi momenta in the range from $\sim 0.2$ fm$^{-1}$ to $\sim 0.5$ fm$^{-1}$. Averaging the effective mass and the quasiparticle residue over the Fermi momentum in the range between $0.2$ fm$^{-1}$ and $0.5$ fm$^{-1}$ we find, respectively, the mean values $m^*_{\downarrow}=1.18\,m$  and $Z_\downarrow=0.78$ using the EM500 interaction, and  $m^*_{\downarrow}=1.19\,m$  and $Z_\downarrow=0.79$, in the case of the AV18 potential. The results for both quantities compare remarkably well with those of the full-many body analysis of Combescot and Giraud \cite{Combescot08} who found $m^*_\downarrow=1.197\,m$, and those of the diagrammatic Monte Carlo method employed by Vlietinck {\it et al.,} \cite{Vlietinck13} who obtained a value of 0.759 for the quasiparticle residue. These results confirm once more the Fermi polaron behavior exhibited by the $\downarrow$ neutron impurity in a low-density free Fermi gas of $\downarrow$ neutrons. 



\section{Summary and Conclusions}
\label{sec:conclusions}

In this work we have reviewed the properties of neutron matter at low-densities. In particular, using the well known BHF approach of nuclear matter and the BCS theory we have calculated, respectively, the ground state energy and the superfluid neutron pairing gap. Results have been obtained for two NN interactions, the chiral one of Entem and Machleidt at N$^3$LO with a 500 MeV cut-off  and the Argonne V18 phenomenological potential. Results for the ground state energy have been also obtained for a simple Gaussian potential with a strength and range adjusted to reproduce the $^1S_0$ neutron-neutron scattering length and effective range. The results have been compared with those of quantum Monte Carlo calculations for neutron matter and cold atoms. We have found that the energy of neutron matter with Fermi momenta in the range $0.35<k_F<0.5$ fm$^{-1}$ is about one half of the energy of a non-interacting Fermi gas, in agreement with previous BHF, variational and finite volume Green's function Monte Carlo calculations of low-density neutron matter. This result indicates that in this range of low densities neutron matter is close to the unitary limit although, actually, it never reaches it. We have determined also the Tan contact parameter in neutron matter finding that
only at very low densities there is a reasonable good agreement  between our results and experimental data form ultra-cold atoms. We have found that out BCS calculation predicts a maximum value of the pairing gap of $\sim 0.5E_F$ for a Fermi momentum of $\sim 0.2$ fm$^{-1}$. However, although this value is close to that found at the unitary limit in experiments ultra-cold Fermi gases, this does not mean that there is a good agreement between our calculation 
and experimental data because the BCS calculation does not take into account medium polarization effects which very important even at low densities and reduce the value of the gap. Finally, we have have reviewed the properties  (energy, effective mass and quasiparticle reside) of a $\downarrow$ neutron impurity in a low-density free Fermi gas of $\uparrow$ neutrons. Our results have shown that this impurity presents properties close to those of an attractive Fermi polaron in the unitary limit.


\section*{Funding}
This work is supported by the COST Action CA16214, ``PHAROS: The multi-messenger physics and astrophysics of neutron stars''.

\section*{Acknowledgements}
The author is very grateful to D. Unkel for very interesting discussions and comments.





\end{document}